\documentclass[a4paper, 10pt, onecolumn]{article}
\pdfoutput=1
\usepackage{graphicx}
\usepackage{multicol}
\usepackage[left=1.7cm,top=1.8cm,right=1.7cm,bottom=1.8cm,nohead]{geometry}
\usepackage{amsmath}

\newcommand{\lesssim}{\, \begin{picture}(8,8)\put(0,2){$<$}\put(0,-3){$\sim$}\end{picture} \,}
\newcommand{\gtrsim}{\, \begin{picture}(8,8)\put(0,2){$>$}\put(0,-3){$\sim$}\end{picture} \,}

\begin{document}
\title{\Large \bf Testing spooky action at a distance}
\author{\normalsize D. Salart, A. Baas, C. Branciard, N. Gisin, and H. Zbinden\\
\it \small Group of Applied Physics, University of Geneva, 20, Rue de l'Ecole de M\'edecine, CH-1211 Geneva 4, Switzerland}
\date{\small \today}
\maketitle

\begin{multicols}{2}
{\bf In science, one observes correlations and invents theoretical models that describe them. In all sciences, besides quantum physics, all
correlations are described by either of two mechanisms. Either a first event influences a second one by sending some information
encoded in bosons or molecules or other physical carriers, depending on the particular science. Or the correlated events have some common
causes in their common past. Interestingly, quantum physics predicts an entirely different kind of cause for some correlations, named
entanglement. This new kind of cause reveals itself, e.g., in correlations that violate Bell inequalities (hence cannot be described by
common causes) between space-like separated events (hence cannot be described by classical communication). Einstein branded it as {\it spooky
action at a distance}.

A real {\it spooky action at a distance} would require a faster than light influence defined in some hypothetical universally privileged reference frame.
Here we put stringent experimental bounds on the speed of all such hypothetical influences. We performed a Bell test during
more than 24 hours between two villages separated by 18 km and approximately east-west oriented, with the source located precisely in the
middle. We continuously observed 2-photon interferences well above the Bell inequality threshold. Taking advantage of the Earth's rotation,
the configuration of our experiment allowed us to determine, for any hypothetically privileged frame, a lower bound for the speed of this spooky influence.
For instance, if such a privileged reference frame exists and is such that the Earth's speed in this frame is less than $10^{-3}$ that of the
speed of light, then the speed of this spooky influence would have to exceed that of light by at least 4 orders of magnitude.}

According to quantum theory, quantum correlations violating Bell inequalities merely happen, somehow from outside space-time, in the sense
that there is no story in space-time that can describe their occurrence: there is not an event here that somehow influences another distant
event there. Yet, such a description of correlations, radically different from all those found in any other part of science, should be
thoroughly tested. And indeed, many Bell tests have already been published\cite{Asp1}. Recently, both the locality\cite{LocLoopholeAspect,LocLoopholeGeneva,LocLoopholeInnsbrug}
and the detection\cite{DetLoopholeRowe,DetLoopholeMat} loopholes have been closed in several independent experiments. Still, one could imagine
that there is indeed a first event that influences the second one. However, the speed of this hypothetical influence would have to be defined in
some universal privileged reference frame and be larger than the speed of light, hence Einstein's condemned it as {\it spooky action at a distance}.
In 1989, Eberhard noticed that the existence of such a hypothetically privileged reference frame could be experimentally tested\cite{Eber}. The idea
is that the speed of this influence, though greater than the speed of light, is finite. Hence, if in the hypothetically privileged frame both
events are simultaneous, then the signal does not arrive on time and no violation of Bell inequalities should be observed. Note that if both
events are simultaneous in a reference frame, then they are also simultaneous with respect to any reference frame moving in a direction
perpendicular to the line joining the two events. Accordingly, Eberhard proposed\cite{Eber2,Scarani} to perform a Bell test over a long distance oriented
east-west during 12 hours. In such a way, if the events are simultaneous in the Earth reference frame, then they are also simultaneous with respect to
all frames moving in the plane perpendicular to the east-west axis and in 12 hours all possible hypothetically privileged frames are scanned.

Bohm's pilot-wave model of quantum mechanics is an example containing an explicit spooky action at a distance\cite{Bohm I}.
As recognized by Bohm, this requires the assumption that there is a universally privileged frame\cite{Bohm II}. In their book\cite{BohmHiley},
Bohm and Hiley also noticed that if the spooky action at a distance propagates at finite speed, then an experiment like the one presented below
could possibly falsify the pilot-wave model. In this book, the authors stress that the existence of a universally privileged frame would not
contradict relativity.

In 2000, some of us already analyzed a Bell experiment along the
lines presented above\cite{Scarani,2000Gisin,2000Zbinden}. However,
the analysis concerned only two hypothetically privileged reference
frames: since that older experiment did not last long enough and was
not oriented east-west, no other reference frame was analyzed. The
first frame was defined by the cosmic background radiation at around
2.7 K. The second frame we analyzed was the "Swiss Alps reference
frame", i.e. not a universal frame, but merely the frame defined by
the massive environment of the experiment. The assumption that the
privileged frame depends on the experiment's environment leads
naturally to question situations where the massive environments on
both sides of the experiment differ, and this was indeed the main
subject of the experiment in 2000\cite{2000Gisin,2000Zbinden}. In
both of these analyses we termed the hypothetical supra-luminal
influence, the {\it speed of quantum information}, to stress that it
is not a classical signaling. We shall keep this terminology, but we
like to emphasize that this is only the speed of a hypothetical
influence and that our result casts very serious doubts on its
existence. Still, it is useful to give names to the objects under
study, even when their existence is hypothetical. For views on the
{\it speed of quantum information}, see\cite{Garisto}.

Before presenting our experiment and results, let us clarify the
principle of our measurements and how one can obtain bounds on this
{\it speed of quantum information} in any reference frame.

In an inertial reference frame centered on the Earth, two events $A$
and $B$ (in our experiment, two single-photon detections) occur at
positions $\vec{r}_A$ and $\vec{r}_B$ at times $t_A$ and $t_B$. Let
us consider another inertial reference frame $F$, the hypothetically
privileged frame, relative to which the Earth frame moves at a speed
$\vec{v}$ (see Figure 1). When correlations violating a Bell
inequality are observed, the {\it speed of quantum information}
$V_{QI}$ in frame $F$ that could cause the correlation is lower
bounded by
\begin{equation}
V_{QI} \geq \frac{||\vec{r'}_B - \vec{r'}_A||}{|t'_B - t'_A|}
\end{equation}
where $(\vec{r'}_A, t'_A)$ and $(\vec{r'}_B, t'_B)$ are the
coordinates of events $A$ and $B$ in frame $F$, obtained from
$(\vec{r}_A, t_A)$ and $(\vec{r}_B, t_B)$ after a Lorentz
transformation. After simplification, one gets
\begin{equation}
\left(\frac{V_{QI}}{c}\right)^2 \geq 1+
\frac{(1-\beta^2)(1-\rho^2)}{(\rho+\beta_{\parallel})^2}
\label{eq_VQI}
\end{equation}
where $\beta = \frac{v}{c}$ is the relative speed of the Earth frame
in frame $F$ ($c$ being the speed of light),
$\beta_{\parallel}=\frac{v_{\parallel}}{c}$, with ${v}_{\parallel}$
the component of $\vec{v}$ parallel to the $AB$ axis, and
$\rho=\frac{c~t_{AB}}{r_{AB}}$ quantifies the alignment of the two
events in the Earth frame (with $t_{AB} = t_B-t_A$ and $r_{AB} =
|\vec{r}_B - \vec{r}_A|$). In the following, we will consider
space-like separated events, for which $|\rho| < 1$: the bound
(\ref{eq_VQI}) on $V_{QI}$ will then be larger than $c$. For a given
privileged frame $F$, this bound depends on the orientation of the
$AB$ axis through $\beta_{\parallel}$ and on the alignment $\rho$.
To obtain a good lower bound for $V_{QI}$, one should upper bound
the term $(\rho+\beta_{\parallel})^2$ by the smallest possible
value, during a period of time $T$ needed to observe a Bell
violation (which, in our experiment, will be the integration time of
a 2-photon interference fringe).

To get an intuition, consider first the simple case where $\rho = 0$
(the two events are perfectly simultaneous in the Earth frame), and
the $AB$ axis is perfectly aligned in the east-west direction. Then,
when the Earth rotates, there will be a moment $t_0$ when the
east-west direction is perpendicular to $\vec v$, i.e.
$\beta_{\parallel}(t_0) = 0$.

\noindent
During a small time interval around
$t_0$, one can bound $|\beta_{\parallel}(t)|$ by a small value, and
thus obtain a high lower bound for $V_{QI}$.

In principle, the alignment $\rho$ could actually be optimized for
each privileged frame that one wishes to test, so as to decrease the
bound that one can put on $(\rho+\beta_{\parallel})^2$ during a time
interval $T$ (and increase the upper term $(1-\rho^2)$ at the same
time). In our experiment, since we want to scan all possible frames,
we will not optimize $\rho$ for each frame; instead, we shall align
the detection events such that $|\rho| \leq \bar{\rho} \ll 1$, where
$\bar{\rho}$ is our experimental precision on the alignment $\rho$.
We shall then use the fact that
$\frac{1-\rho^2}{(\rho+\beta_{\parallel})^2} \geq
\frac{1-\bar{\rho}^2}{(\bar{\rho}+|\beta_{\parallel}|)^2}$, to get
the bound
\begin{equation} \left(\frac{V_{QI}}{c}\right)^2 \geq 1+
\frac{(1-\beta^2)(1-\bar{\rho}^2)}{(\bar{\rho}+|\beta_{\parallel}|)^2}
\label{eq_VQI_bis}
\end{equation}
The problem reduces to bounding $|\beta_{\parallel}|$ directly.
\begin{center}
\includegraphics[width=1.0\linewidth]{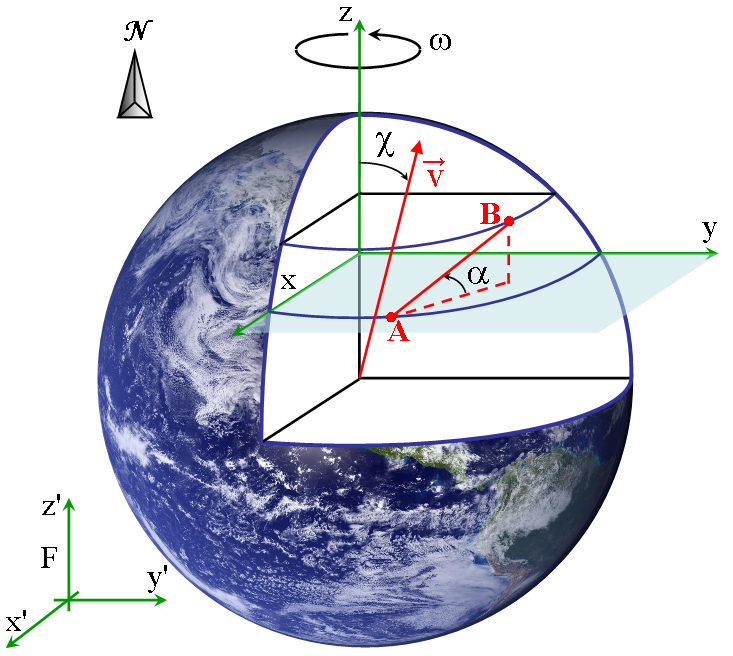}
\end{center}
{\small {\bf Figure 1 $\mid$ Reference frames.} The Earth frame
moves with respect to a hypothetically privileged reference frame
$F$ at a speed $\vec{v}$. The zenith angle $\chi$ between $\vec{v}$
and the $z$ axis can have values between $0^\circ$ and $180^\circ$.
The $AB$ axis forms an angle $\alpha$ with the equatorial ($xy$) plane. $\omega$ is the angular velocity of the Earth.}\\

In the configuration of our experiment, the $AB$ axis is almost, but
not perfectly, oriented along the east-west direction. Consequently,
the component $\beta_{\parallel}(t)$ has a 24-hour period, and
geometric considerations show that it can be written as (see the
Methods section)
\begin{equation}
\beta_{\parallel}(t) = \beta \cos \chi \sin \alpha + \beta \sin \chi
\cos \alpha \cos \omega t \ ,
\end{equation}
where $\chi$ is the zenith angle of $\vec v$, $\alpha$ is the angle
between the $AB$ axis and the equatorial ($xy$) plane (see Figure
1), and $\omega$ is the angular velocity of the Earth.

As we show in the Methods section, in order to upper bound
$|\beta_{\parallel}|$ during a period of time $T$, one can consider
two cases, depending whether $\vec v$ points close to a pole or not:
\begin{eqnarray}
(i) && C_T \, |\tan \chi| > |\tan \alpha| \label{case_i} \\
(ii) && C_T \, |\tan \chi| \leq |\tan \alpha| \ \label{case_ii}
\end{eqnarray}
with $C_T = \cos^2 \frac{\omega T}{4} \simeq 1$ when $\omega T$ is
small. For each case, there exists a time interval of length $T$,
during which $|\beta_{\parallel}(t)|$ is respectively upper-bounded
by
\begin{eqnarray}
(i) && \!\!\!\!\!\! |\beta| \ \sqrt{\sin^2 \chi
\cos^2 \alpha - \cos^2 \chi \sin^2 \alpha} \ \frac{\omega T}{2} \\
(ii) && \!\!\!\!\!\! |\beta| \ \Big( |\cos \chi \sin \alpha| - |\sin
\chi \cos \alpha| \cos \frac{\omega T}{2} \Big) \label{bound_ii}
\end{eqnarray}
These bounds, together with equation (\ref{eq_VQI_bis}), provide the
desired lower bound for $V_{QI}$.

We now describe our experiment. In essence it is a large Franson interferometer\cite{Franson}. A source situated in our laboratory in down-town
Geneva emits entangled photon pairs using the standard parametric down-conversion process in a nonlinear crystal (here a cw laser pumps a waveguide in a
Periodically Poled Lithium Niobate (PPLN) crystal)\cite{SebEPJD}. Using fiber Bragg gratings and optical circulators, each pair is deterministically split
and one photon is sent via the Swisscom fiber optic network to Satigny, a village west of Geneva, while the other photon is sent to Jussy, another
village east of Geneva. The two receiving stations, located in those two villages, are separated by a direct distance of 18.0 km, see Figure 2. We use
energy-time entanglement, a form of entanglement well suited for quantum communication in standard telecom fibers\cite{Rob}. At each receiving station,
the photons pass through identically unbalanced fiber optic Michelson interferometers. The imbalance ($\approx 25$ cm) is larger than the single-photon
coherence length ($\approx 2.5$ mm), hence avoiding any single-photon interference, but much smaller than the pump laser coherence length ($>20$ m).
Accordingly, when a photon pair is detected simultaneously in Satigny and Jussy, there is no information about which path the photons took in their
interferometer, the long arm or the short arm. But since both photons were also emitted simultaneously, both took the same path: both long or both short.
This indistinguishability leads, as always in quantum physics, to interference between the long-long and short-short paths. By scanning continuously the
phase in one interferometer, at Jussy, while keeping the other one stable, produces a sinusoidal oscillation of the correlation between the photon
detections at Satigny and Jussy.

During each run of the experiment we continuously monitored both the single count rates (as a check of the stability of the entire setup) and the
coincidence count rate. The average coincidence rate was 33 coinc./min. and the number of accidental coincidences 2.5 coinc./min. We are primarily
interested in the coincidence rate as its oscillations follow the scanned phase and should have a visibility large enough to exclude any common
cause explanation. The correlations are thus either due to entanglement, as predicted by quantum physics, or due to some hypothetical
{\it spooky action at a distance} whose speed we wish to lower bound.

The phases are controlled by the temperature of the fiber-based interferometers.
To scan the temperature in one of the interferometers, a voltage ramp is applied to its temperature controller. The temperature
decreases regularly for several hours and is then heated quickly at the end of the ramp. This process was repeated during several days.
The end of the cooling ramp stops the phase scan for several minutes making impossible to obtain arbitrarily long measurements of uninterrupted
fringes.

Figure 3 presents a measurement run over 4 hours with a fringe period of $T$=900 s and a continuous sinusoidal fit. This result is remarkable
because the period of the interference fringes remains stable for a very long time, wich allows us to fit the entire measurement with a continuous
fit and obtain a high visibility value.

\end{multicols}
\begin{center}
\includegraphics[width=0.97\linewidth]{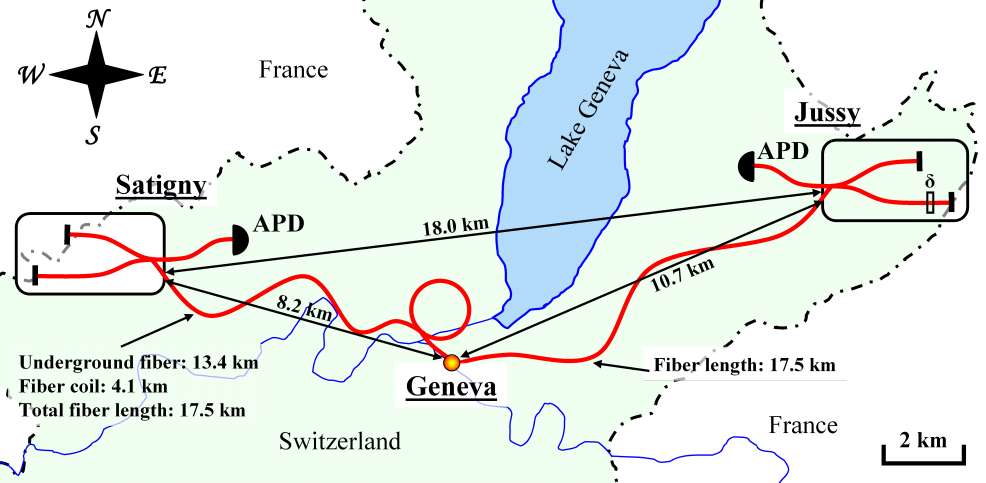}
\end{center}
{\small {\bf Figure 2 $\mid$ Experimental setup.} The source sends pairs of photons from Geneva to two receiving stations through the Swisscom fiber
optic network. The stations are situated in two villages (Satigny and Jussy) in the Geneva region, at 8.2 and 10.7 km, respectively. The direct
distance between them is 18.0 km. At each receiving station, the photons pass through identically unbalanced Michelson interferometers and are
detected by a single-photon InGaAs APD (id Quantique, id201). The length of the fiber going to Jussy is 17.5 km. The fiber going to Satigny was only
13.4 km long so we added a fiber coil of 4.1 km (represented as a loop) to equalize the length in both fibers. Having fibers with the same length allows
us to satisfy the condition of good alignment ($\rho\ll1$).}
\begin{multicols}{2}

The bound for $V_{QI}$ is higher for shorter fringe periods $T$. To reduce the time $T$, one should increase the rate of the
phase scan (more degrees per unit of time). Unfortunately, with a higher rate, the number of coincidences per minute diminishes, hence, the slope
of the temperature ramp was adjusted so as to obtain a compromise period of $T=360$ seconds.
\begin{center}
\includegraphics[width=1.0\linewidth]{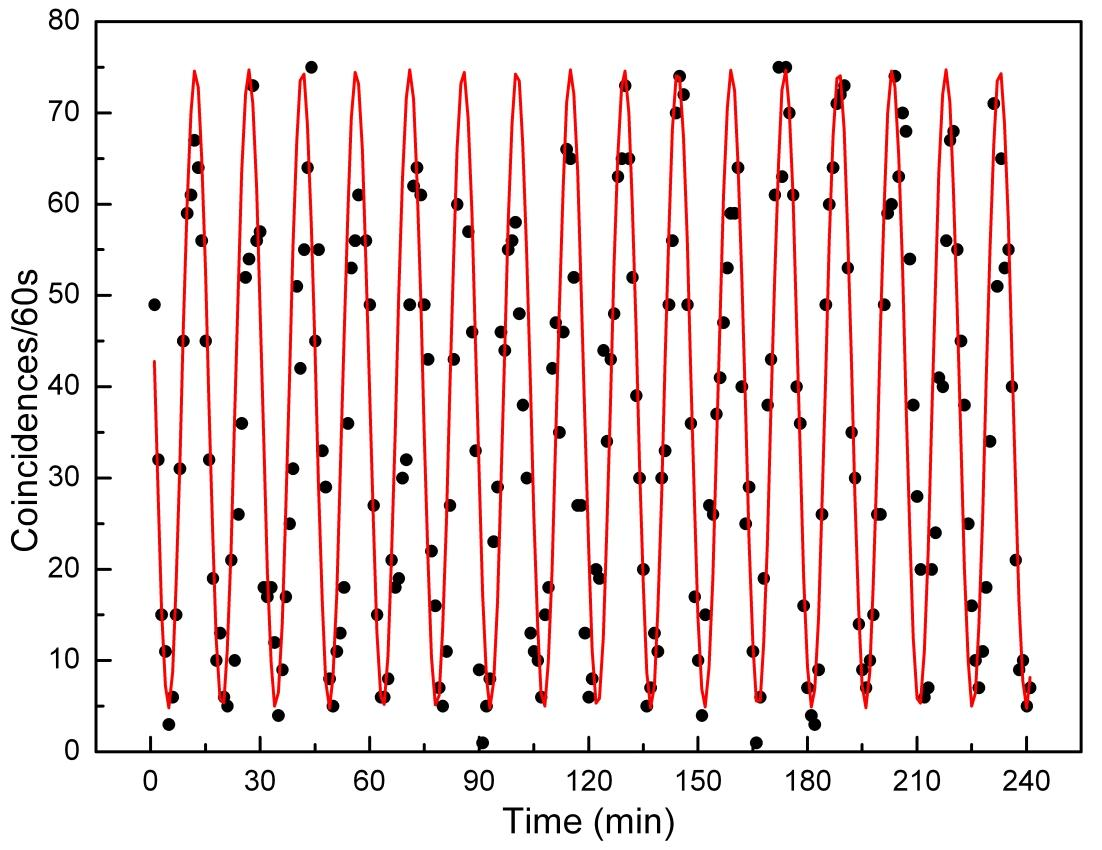}
\end{center}
{\small {\bf Figure 3 $\mid$ } Interference fringes with a period $T$=900 s obtained during a 4 hours measurement fitted with a sinusoidal function
yielding a visibility of $V=(87.6\pm 1.1)\%$. If we substract the accidental coincidences, the visibility climbs to $V_{net}=(94.1\pm 1.0)\%$.}\\

Interference fringes were recorded in many runs usually lasting several hours, up to 15 hours for the longest run. The limitations for the
length of these measurements were the end of the cooling ramp and small instabilities in the setup that produced short interruptions in the scan.
Juxtaposing several of these measurement runs obtained over several weeks, we covered a 24-hour period with interference fringes periods of $T$=360 s
with visibilities well above the threshold ($V=\frac{1}{\sqrt{2}}$) set by the CHSH Bell inequality\cite{CHSH}.

Since long measurements of fringes with short fringe periods $T$ are difficult to fit continuously, we fit the data over a time-window corresponding to
one and a half fringe and scanned this time-window, as explained in the Supplementary information. The results are presented in Figure 4.
\begin{center}
\includegraphics[width=1.0\linewidth]{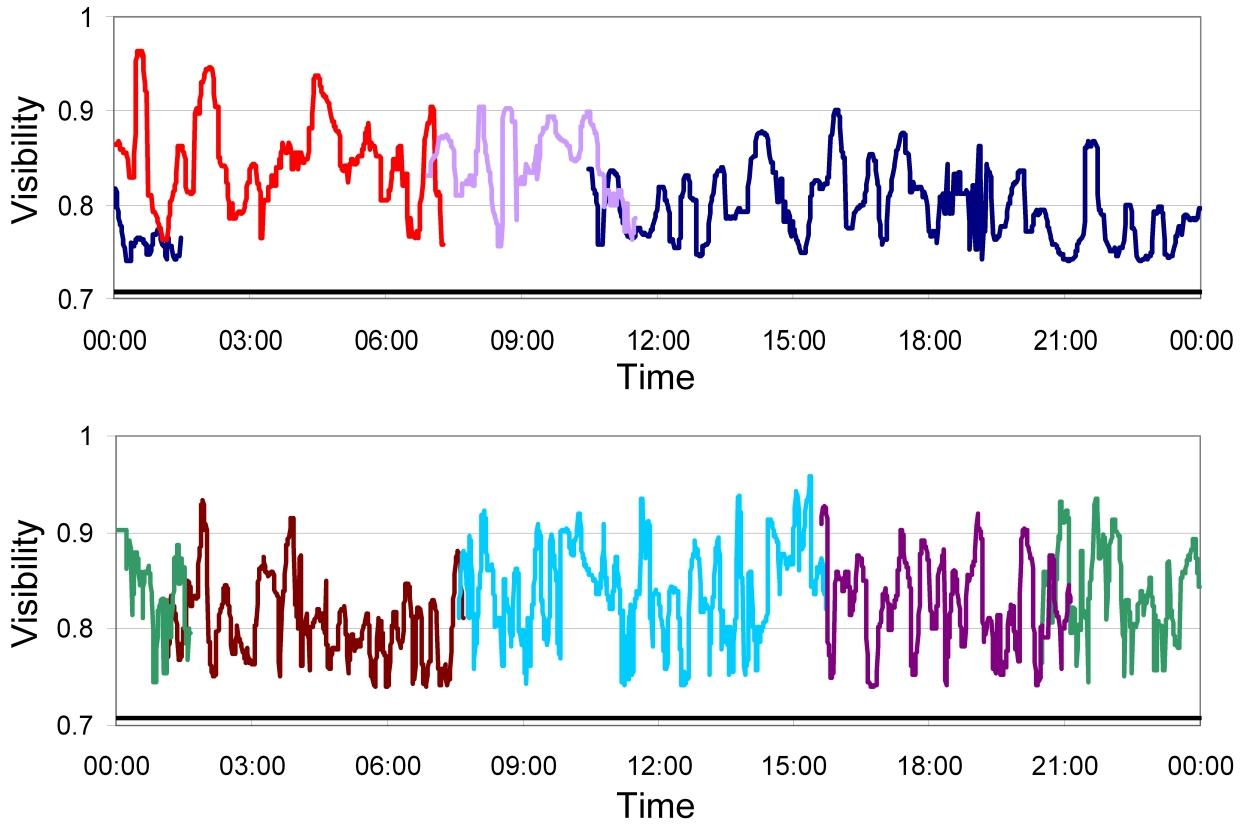}
\end{center}
{\small {\bf Figure 4 $\mid$ }
Visibility fits for several uninterrupted runs obtained at different times of the day were collected. Together these runs cover
each moment of the day at least twice. Visibility values remain above the threshold (black line, $V=\frac{1}{\sqrt{2}}$) set by
the CHSH Bell inequality at all times.}\\

The violation of the Bell inequality at all times of the day allows
one to calculate the lower bound for the {\it speed of quantum
information} for any reference frame. This bound depends on
precision of the alignment in the actual experiment, see eq. (3).
Since we wish to have a good alignment ($\rho\ll1$),
the difference in the arrival times of the single-photons ($t_{AB}$)
should be minimized. First, the length of each fiber between the
source and the single-photon detectors was measured. The long fibers
(of several km) were measured using a single-photon Optical Time
Domain Reflectometer ($\nu$-OTDR)\cite{nu-OTDR} and the short fibers
(less than 500 m) were measured with an Optical Frequency Domain
Reflectometer (OFDR)\cite{OFDR}. The fiber on the Satigny side was
found to be shorter by 4.1 km. We added a fiber coil to the short
side (represented as a loop in figure 2), reducing this difference
to below one centimeter with an uncertainty of 1 cm which
corresponds to a time of 49 ps. To avoid controversy over where
exactly the measurement takes place, we adjusted the fiber lengths
from the source to the fiber couplers inside each interferometer and
also to the photodiodes (where the photons are detected). Hence the
configuration is totally symmetric. Next, we considered the
chromatic dispersion in the fibers. Chromatic dispersion adds an
uncertainty in the arrival times, and because the entangled photons
are anticorrelated in energy, their time delay is always opposite to
each other, thus always increasing this uncertainty. Chromatic
dispersion was measured to be 18.2
$\frac{\textrm{ps}}{\textrm{nm}\cdot \textrm{km}}$ using a Chromatic
Dispersion Analyzer\cite{CDA}. For a spectral half width of
$\Delta\lambda = 0.5$ nm and two times the distance of 17.55 km,
this is equivalent to 319 ps of uncertainty. Thus, the overall
uncertainty in the relative lengths of the fibers is $t_{AB}=323$
ps. This, together with the direct distance between the receivers,
$r_{AB}=18.0$ km, allows one to estimate the precision of our
alignment: $|\rho| \leq \bar{\rho} = 5.4\cdot10^{-6}\ll1$. 

Finally, we use equation $(3)$ to calculate a lower bound for
$V_{QI}$. We use the value of $\bar{\rho}$ just calculated, the
period of time $T=360$ s needed to observe a Bell violation
(corresponding to the interference fringe period) and the angle
formed by the axis between the two receiving stations (axis $AB$)
and the east-west direction, $\alpha=5.8^\circ$. The results are
shown in Figures 5a and 5b, for certain hypothetically privileged
frames. In Figure 5a, we scan all possible directions $\chi$,
but set the Earth's relative speed at $\beta=10^{-3}$.
A lower bound for $V_{QI}$ greater than $10.000$
times the speed of light is found for any such reference frame. The
non-perfect east-west orientation ($\alpha\ne0$) is responsible for
the minimum values of the bound at angles $\chi$ near $0^\circ$ and
$180^\circ$. For smaller Earth speeds, the bound on $V_{QI}$ is even
larger. On the other hand, if $\beta$ is very large, then the
corresponding bound on $V_{QI}$ is less stringent. To illustrate
this, in Figure 5b we set $\chi=90^\circ$, i.e. $\vec v$ in the
equatorial plane, and scan the velocity $\beta$.
Indeed, when $\beta \simeq 1$, the bound drops rapidly.
Recall, however, that for large values of $\beta$ one could, in principle,
optimize the alignment $\rho$ in the experiment, so as to get a better
bound on $V_{QI}$. For small values of $\beta$, our bound is limited by
the inverse of our precision of alignment $\bar{\rho}$.
\end{multicols}
\begin{center}
\includegraphics[width=0.49\linewidth]{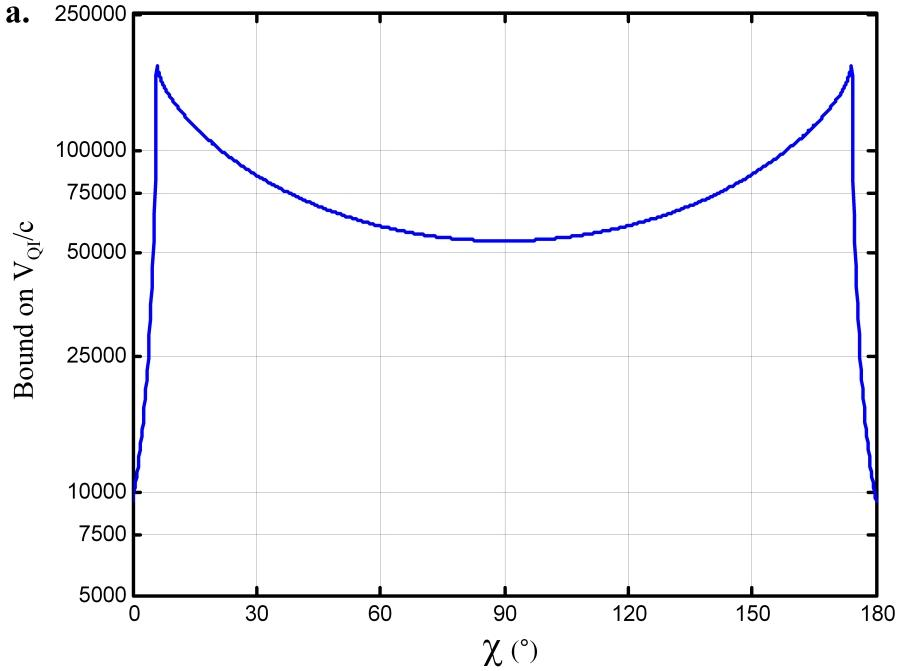}
\includegraphics[width=0.49\linewidth]{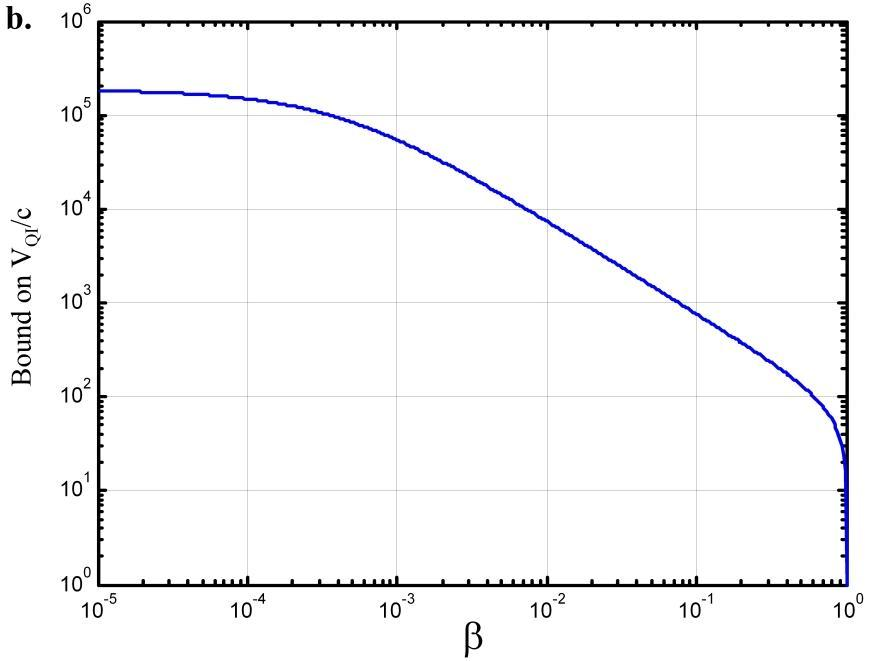}
\end{center}
{\small {\bf Figure 5 $\mid$ Lower bounds for the {\it speed of
quantum information}. a.} Bound obtained for $\frac{V_{QI}}{c}$ as a
function of the angle $\chi$, when $\beta=10^{-3}$. For angles $\chi
\lesssim \alpha$ or $\chi \gtrsim 180^\circ-\alpha$, the bound is
obtained by considering case $(ii)$ (see text), while for angles
$\alpha \lesssim \chi \lesssim 180^\circ-\alpha$, the bound is
obtained by considering case $(i)$. The bound at $\chi=90^\circ$ is
$V_{QI} \geq 54000 c$. {\bf b.} Bound obtained for
$\frac{V_{QI}}{c}$ as a
function of the speed $\beta$, when $\chi=90^\circ$. When $\beta \rightarrow 0$, our bound on $\frac{V_{QI}}{c} \rightarrow 1/\bar{\rho}$.}\\
\begin{multicols}{2}
In conclusion, we performed a Bell experiment using entangled photons between two villages separated by 18 km and approximately east-west
oriented, with the source located precisely in the middle. The rotation of the Earth allowed us to test all possible hypothetically privileged
frames in a period of 24 hours. Two-photon interferences fringes with visibilities well above the threshold set by the Bell inequality
were observed at all times of the day.

From these observations we conclude that the nonlocal correlations observed here and in previous experiments\cite{Asp1} are indeed truly nonlocal.
Indeed, to maintain an explanation based on {\it spooky action at a distance} one would have to assume that the spooky action propagates at speeds
even greater than the bounds obtained in our experiment.

{\small {\bf Acknowledgements} We acknowledge technical support by J-D. Gautier and C. Barreiro. The access to the telecommunication network
was provided by Swisscom. This work was supported by the Swiss NCCR Quantum Photonics and the EU project QAP. Credit for the image of the Earth
in figure 1 goes to NASA Goddard Space Flight Center Image by Reto St\"ockli.}

\end{multicols}

\end{document}